\documentclass{article}

\usepackage{PRIMEarxiv}
\usepackage{amsmath}
\usepackage{multirow}
\usepackage[utf8]{inputenc} 
\usepackage[T1]{fontenc}    
\usepackage{hyperref}       
\usepackage{url}            
\usepackage{booktabs}       
\usepackage{amsfonts}       
\usepackage{nicefrac}       
\usepackage{microtype}      
\usepackage{lipsum}
\usepackage{fancyhdr}       
\usepackage{graphicx}       
\graphicspath{{media/}}     

\title{Cross-Subject EEG Emotion Recognition Based on Temporal Asynchronous Alignment Contrastive Learning}

\author{
\textnormal{
Ying Xie, Yi Zheng, Zehui Xiao, Wenkai Lu, Mengting Liu\thanks{
Corresponding author: Mengting Liu. \\
Y. Xie, Y. Zheng, Z. Xiao and M. Liu are with the School of Biomedical Engineering, Shenzhen Campus of Sun Yat-sen University, Shenzhen 518107, China 
(e-mail: \{xiey285, zhengy323, xiaozh26\}@mail2.sysu.edu.cn; liumt55@mail.sysu.edu.cn). \\
W. Lu is with the School of Computer Science and Technology, Tianjin University, Tianjin 300354, China 
(e-mail: 3020244261@tju.edu.cn).
}
}
}

\begin{document}
\maketitle

\begin{abstract}
With the advancement of science and technology, the importance of emotion research has become increasingly evident. Electroencephalography (EEG)-based emotion recognition has emerged as an active research area in recent years, owing to its objectivity and high temporal resolution. However, most existing methods focus on optimizing encoder structures to enhance feature extraction capabilities, while paying relatively little attention to similarity calculation strategies, particularly overlooking the potential temporal misalignment of responses among different subjects. To address these shortcomings, this paper draws inspiration from the late interaction mechanism of ColBERT in natural language processing (NLP) and proposes a Temporal Asynchronous Alignment-based Contrastive Learning (TA2CL) framework. This method transforms the traditional global "hard alignment" similarity calculation approach into a fine-grained local matching mechanism, enabling the model to adaptively search for and align "locally highly correlated" segments between two EEG signals, thereby effectively mitigating the effects of inter-subject differences and temporal delays. Experimental results demonstrate that the proposed method achieves strong performance across multiple public datasets. Specifically, on the FACED dataset, it achieves an accuracy of 64.5\% for the nine-class classification task and 79.5\% for the binary classification task, while on the SEED and SEED-V datasets, it achieves accuracies of 86.4\% and 70.1\%, respectively, validating the method's effectiveness and generalization capability.
\end{abstract}


\begin{figure}[t]
\centering
\includegraphics[width=0.8\columnwidth]{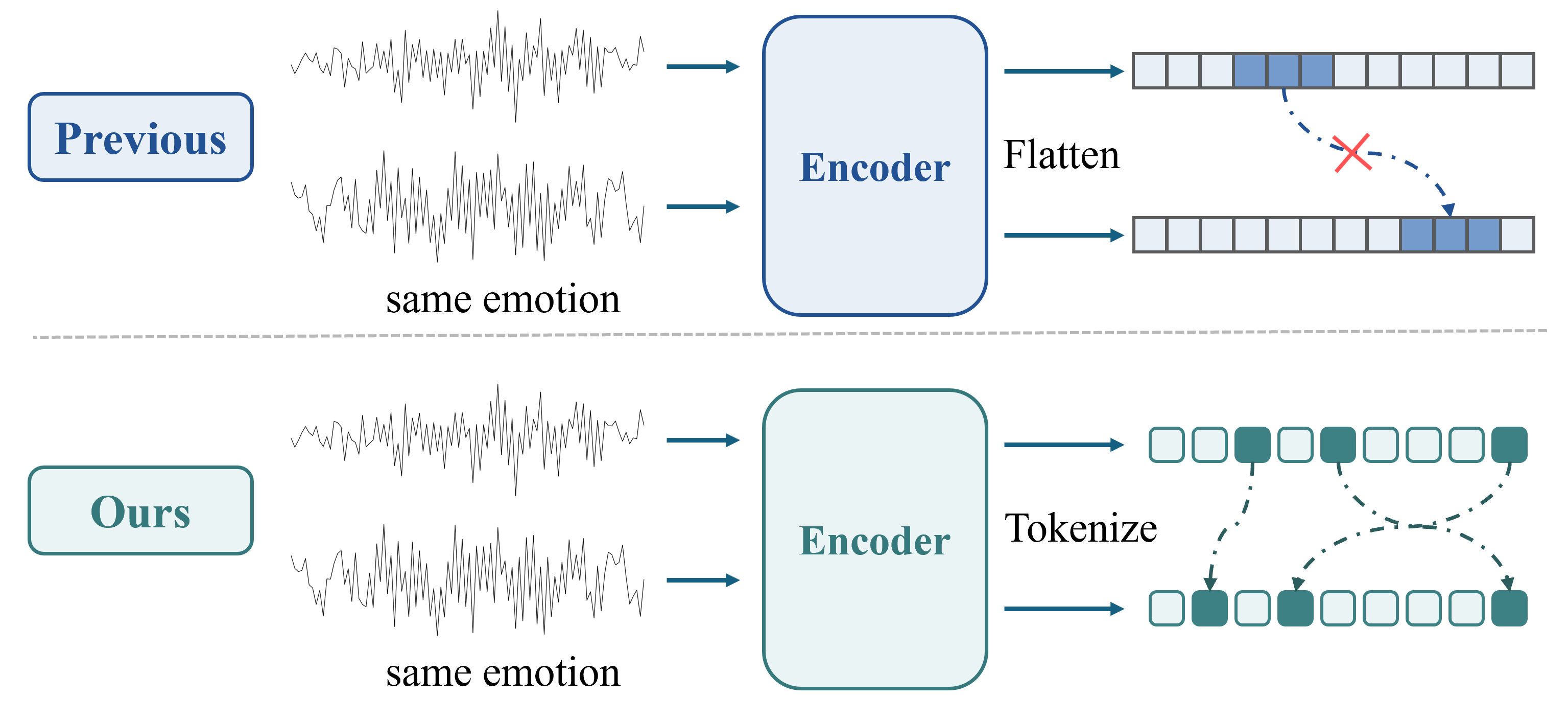}
\caption{Comparison of our method with previous methods. Previous methods calculate sample similarity based on global features, which are susceptible to temporal asynchrony in EEG signals, making it difficult to match local similar patterns within samples of the same emotion. Our method retains the encoded temporal token representations and performs local matching between sequences through asynchronous contrastive learning. This allows similar emotional segments to contribute high positive sample similarity even when they appear at different temporal positions, leading to more robust emotional representations.}
\label{fig:introduction}
\end{figure}

\section{Introduction}
\label{sec:introduction}
Emotions are a key manifestation of human psychological states. They are not only closely linked to an individual's cognitive activities and communication decisions but also play an important role in physical and mental health \cite{picard2003affective}. As societal concern regarding mental health continues to grow, particularly in light of the increasing prevalence of affective disorders such as depression and anxiety, objective emotion quantification and detection have emerged as an active research area, attracting increasing attention from scholars \cite{garcia2018mental}. 

In recent years, with the rapid development of artificial intelligence and affective computing technologies, deep learning has been widely applied in the field of emotion recognition due to its powerful data processing capabilities and advantages in nonlinear feature extraction \cite{craik2019deep}. Researchers typically employ Convolutional Neural Networks (CNNs) to extract spatial features from multi-channel signals \cite{kim2022accelerating,jadhav2023eeg}, or utilize Recurrent Neural Networks (RNNs) and their variants to model the temporal dynamics of signals \cite{yang2018emotion,chen2015multi}. However, research indicates that when humans perform complex tasks or face changes in the external environment, their internal brain networks undergo real-time reorganization and specialization. Negative emotions or physiological states, such as situational stress, can significantly disrupt the adaptive transition from a resting state to a task state, and may even affect an individual's executive control performance by inhibiting the brain's adaptive reorganization \cite{liu2021context,liu2021brain}. Therefore, effectively capturing connectivity patterns among brain networks is crucial for improving the performance of emotion recognition.
Meanwhile, when graph neural networks (GNNs) are used to model complex whole-brain connectivity, effectively addressing the computational challenges posed by high node density while capturing the spatial heterogeneity of different brain regions and maintaining model interpretability remains a bottleneck in computational neuroscience \cite{li2025surfgnn}.
Among the various emotion recognition methods, non-physiological signals such as facial expressions and speech, while easily obtainable, are susceptible to individual subjective control and may be intentionally masked or concealed; furthermore, differences in interpretation across cultural backgrounds can also affect recognition results \cite{gamage2022driver}. In contrast, physiological signals can reflect an individual's true emotional state in a more stable and objective manner \cite{alarcao2017emotions}. Among these, EEG-based emotion recognition methods offer advantages such as non-invasiveness, low cost, high temporal resolution, and rich information content. Capable of capturing emotional changes in the brain on a millisecond timescale, they have gradually become a major research direction at the intersection of artificial intelligence and brain-computer interfaces \cite{suhaimi2020eeg,wang2023deep}.

In recent years, contrastive learning has achieved significant progress in cross-subject EEG emotion recognition, demonstrating considerable application potential \cite{mohsenvand2020contrastive,shen2022contrastive}. Existing contrastive learning-based studies commonly adopt a three-stage "Encoder–Projector–Classifier" architecture \cite{chen2020simple}. To address the inconsistent signal distributions across individuals caused by physiological differences, most studies have focused primarily on optimizing the encoder structure, aiming to better capture inter-subject variability and learn more refined and discriminative emotion representations.
For example, Lin et al. \cite{lin2025spatial} proposed a dual-branch encoding architecture based on a spatiotemporal Transformer. This method introduces an independent graph-space encoder to characterize the dynamic topological relationships among electrode channels, and combines it with a temporal encoder equipped with attention mechanisms to extract long-range dependencies across temporal scales. Shen et al. \cite{shen2025dynamic} proposed DAEST, which incorporates a dynamic attention mechanism to model spatiotemporal variations in EEG signals. By adaptively weighting the importance of different time segments and electrode channels, DAEST highlights emotion-related key features while suppressing interference from redundant or irrelevant information, thereby enhancing feature representation.

Although these methods improve recognition accuracy and classification performance to some extent through encoder optimization, most of them overlook another critical aspect of contrastive learning: the design of similarity computation strategies. In existing frameworks, feature similarity is typically measured using cosine similarity after global features are concatenated and transformed into one-dimensional vector representations \cite{chen2020simple}. Such coarse-grained global feature matching may weaken the dynamic characteristics and fine-grained temporal information of EEG signals \cite{eldele2021time}. As a result, it becomes difficult to effectively address the temporal asynchrony of emotional responses across individuals \cite{mohsenvand2020contrastive}, which further limits the improvement of cross-subject generalization ability.

To address these limitations, this paper moves beyond traditional global feature matching and draws inspiration from late interaction mechanisms in natural language processing. In conventional text retrieval and matching tasks, models typically compress an entire sentence into a single global feature vector. However, such coarse-grained representations may lose critical information and semantic details \cite{reimers2019sentence,humeau2019poly}. To overcome this issue, the ColBERT architecture \cite{khattab2020colbert} introduces a fine-grained matching mechanism based on late interaction. Specifically, it preserves multidimensional feature representations at the token level rather than compressing each sequence into a single global vector. During similarity computation, ColBERT computes the similarity between each token in the query sequence and all tokens in the target sequence, and then applies the MaxSim operator to select the maximum similarity response for each query token. These maximum responses are further aggregated to obtain the final sequence-level similarity score. This strategy enables precise alignment and interaction between sequences at the level of local details.

EEG signals and sequence data in NLP share certain structural similarities, as both can be regarded as dynamic sequences containing rich contextual information. Consequently, deep learning models such as Transformers have been widely applied in both fields \cite{song2022eeg}. However, existing classical self-supervised contrastive learning frameworks, such as SimCLR \cite{chen2020simple}, typically rely on global features to perform strict "hard alignment" when processing EEG data. Specifically, EEG features are often flattened or projected into one-dimensional vector representations, and their similarity is then measured using cosine similarity. Such global representation-based matching tends to weaken the dynamic evolution of signals along the temporal dimension, leading to the loss of fine-grained temporal information \cite{eldele2021time}. Moreover, due to individual differences in physiological structures and cognitive patterns, EEG responses elicited by the same emotional stimulus may exhibit latency variations across subjects. As a result, emotion-relevant segments may appear asynchronously along the temporal axis \cite{mariooryad2013analysis}. Under such circumstances, forced alignment based on global representations struggles to effectively model cross-temporal delays, thereby limiting the model's ability to capture subject-invariant emotion-related features. Based on the above analysis, this paper introduces and reconstructs the MaxSim mechanism from ColBERT and extends it to EEG temporal modeling. Accordingly, we propose Async-InfoNCE, a contrastive learning loss function tailored to inter-subject temporal asynchrony. As shown in Figure \ref{fig:introduction}, the proposed method performs fine-grained feature matching at the local time-slice level, effectively mitigating the information loss caused by traditional global alignment strategies and enhancing the robustness and generalization ability of cross-subject emotion recognition.
Specifically, Async-InfoNCE no longer forces multidimensional features to be compressed into a single global representation vector, but instead preserves the temporal information of EEG signals in the representation space \cite{franceschi2019unsupervised}. During the contrastive learning optimization process, this method transforms the original coarse-grained global "hard alignment" mechanism into a more flexible fine-grained "soft matching" strategy \cite{cuturi2017soft}, enabling the model to overcome the constraints of absolute temporal alignment and adaptively search for and match "local high correlations" across multiple time slices of the two EEG signal segments. This asynchronous alignment mechanism not only effectively suppresses noise interference in individual neural responses but also captures stable and consistent deep emotional representations across subjects at a finer temporal scale, thereby enhancing the model's ability to distinguish emotional features and its generalization performance.

Based on the above innovations, the main contributions of this paper can be summarized as follows:
\begin{itemize}
\item We propose TA2CL, a contrastive learning framework for cross-subject EEG emotion recognition. For the first time, we introduce late interaction mechanisms from NLP into EEG contrastive learning tasks. By constructing an asynchronous feature alignment strategy based on Async-InfoNCE, we achieve fine-grained modeling of cross-subject emotional features, providing a novel approach to address the issue of individual heterogeneity in EEG signals.
\item Building upon the traditional MaxSim similarity calculation mechanism, we design an improved strategy that combines ${TopK}$ matching with mean aggregation. By introducing multiple local matches, this method effectively avoids information loss caused by a single optimal match. Additionally, scale control is introduced to improve the numerical stability of similarity computation, enabling the model to better capture the temporal dynamics of EEG signals.
\item Extensive experimental validation was conducted on multiple public datasets (SEED, SEED-V, and FACED). The results demonstrate that the proposed method achieves competitive performance across various emotion classification tasks, enhances the model's cross-subject generalization capability, and validates the effectiveness of the proposed alignment mechanism in capturing stable emotional features.
\end{itemize}

\section{Related Work}
\label{sec:related}

\subsection{Cross-Subject EEG Emotion Recognition Based on Contrastive Learning} 

Due to inter-subject differences in physiological structures and cognitive patterns, EEG signals often exhibit significant distribution shifts in cross-subject emotion recognition. To mitigate this issue, early studies commonly adopted transfer learning or unsupervised domain adaptation methods. For instance, Zheng et al. \cite{zheng2016personalizing} explored several classical transfer learning strategies to mitigate inter-subject distribution shifts and facilitate personalized adaptation in EEG-based emotion recognition. Li et al. \cite{li2019domain} employed an adversarial domain adaptation framework to learn more consistent latent representations between source and target domains, effectively alleviating distribution discrepancies in cross-subject settings. Nevertheless, these methods generally require target-domain data during training. Although they can improve recognition performance for specific target subjects, their generalization ability is still limited when applied to unseen subjects or unknown data distributions, which restricts their practical applicability \cite{kim2025domain}.

To reduce reliance on target-domain data, contrastive learning has recently been widely explored in EEG-based emotion recognition, as it can learn subject-invariant representations with strong generalization ability without requiring explicit target-domain adaptation \cite{wen2024transformer,wang2023supervised}. For instance, Mohsenvand et al. \cite{mohsenvand2020contrastive}introduced the classical SimCLR framework into EEG time-series modeling. By designing EEG-specific augmentation strategies, including temporal perturbation, amplitude scaling, and frequency-domain filtering, as well as channel reorganization mechanisms, their method reduces the representation distance between semantically similar signal segments and learns more discriminative features. Shen et al. \cite{shen2022contrastive} later proposed CLISA, which employs contrastive learning to pull similar samples closer while pushing dissimilar samples apart. By maximizing the spatiotemporal similarity of EEG samples associated with the same emotional states across subjects, CLISA effectively improves cross-subject generalization. In addition, several studies have integrated contrastive learning with graph neural networks to capture more complex spatial relationships and structural dependencies in EEG signals within a self-supervised learning framework, further enhancing representation learning capability \cite{li2022gmss}.

Although the aforementioned contrastive learning-based methods have made considerable progress in cross-subject EEG emotion recognition, most existing studies primarily focus on optimizing front-end feature encoders to extract more discriminative representations. However, in contrastive loss computation, they generally still rely on basic similarity metrics constructed from simple feature concatenation and cosine similarity. This coarse-grained global alignment strategy tends to flatten the dynamic temporal evolution of EEG signals and neglects the potential temporal asynchrony across subjects during emotion elicitation. As a result, the model struggles to capture finer-grained temporal dependencies, which becomes a key bottleneck limiting further improvements in emotional representation learning under contrastive learning frameworks.

\subsection{Fine-grained alignment of time series and late interaction mechanisms} 

Fine-grained feature alignment methods have demonstrated strong performance in both NLP and time-series analysis \cite{zhou2023one}. In NLP, to address information incompleteness and semantic information loss caused by compressing long sequences into a single global vector, researchers have proposed modeling frameworks based on late interaction. For example, Khattab et al. \cite{khattab2020colbert} proposed ColBERT, a BERT-based architecture that preserves multidimensional token-level representations and introduces the MaxSim similarity computation mechanism to achieve fine-grained matching between sequences. In time-series analysis, to address data distortion and temporal shifts along the timeline, dynamic time warping methods based on elastic alignment have been proposed. For instance, Soft Dynamic Time Warping (Soft-DTW) \cite{cuturi2017soft} establishes flexible matching relationships between different time steps, thereby improving the model's robustness to temporal misalignment. These studies suggest that fine-grained alignment mechanisms can effectively preserve local sequential information and provide a useful basis for modeling temporal asynchrony in EEG signals.

The evolution from "hard alignment", driven by global pooling, to "soft alignment" based on local feature matching is considered a significant trend in complex sequence comparison learning \cite{cui2025physiosync}. However, relevant research in the field of EEG-based affective computing remains relatively scarce. EEG signals are characterized by significantly high temporal resolution and dynamic variability, while latency variations are commonly observed across individuals during emotion induction \cite{brouwer2012estimating}. In cross-subject emotion recognition scenarios, there remains a lack of systematic research and in-depth discussion on how to introduce fine-grained asynchronous matching mechanisms into this field and further integrate them into the design of loss functions for self-supervised contrastive learning.

\section{Method}
We proposes a contrastive learning framework based on fine-grained temporal asynchronous alignment (TA2CL) and introduces a novel contrastive loss function, Async-InfoNCE, to overcome the limitations of global-feature-based coarse temporal alignment in existing methods, thereby enabling the model to capture more discriminative emotion-related representations at fine-grained temporal scales.

\subsection{Definition of Cross-Subject Issues and Spatio-Temporal Feature Encoding}

\subsubsection{Definition of the Cross-Subject EEG Recognition Problem}
This paper aims to construct a cross-subject EEG emotion recognition model with strong generalization ability. Following the setting of existing contrastive learning frameworks \cite{shen2022contrastive}, the data from subjects involved in training are defined as the source domain, denoted as $D_{Source}=\left\{X_i,y_i\right\}_{i=1}^{M_{Soucse}}$, while the data from unknown subjects not involved in training are defined as the target domain, denoted as $D_{Target}=\left\{X_i,y_i\right\}_{i=1}^{M_{Target}}$. Since EEG signals typically exhibit significant distribution shifts between the two domains, the core objective of this paper is to learn subject-independent and generalizable emotion-related spatiotemporal representations from the source-domain data via contrastive learning, without relying on target-domain data for adaptation, thereby improving the model's generalization performance on unseen subjects.

\subsubsection{An emotion-based cross-subject sampling strategy}
In contrastive learning frameworks, the construction of positive and negative samples plays a crucial role in determining model performance. Traditional EEG contrastive learning methods often rely on data augmentation strategies, such as random noise injection, to generate positive sample pairs. In contrast, following the ISC-based positive-pair construction strategy introduced by the CLISA framework \cite{shen2022contrastive}, this paper assumes that when different subjects are exposed to the same external stimulus, such as watching the same emotional video, similar neural activity patterns may be elicited across subjects. Based on this theory, this paper adopts a stimulus-alignment-based sample construction strategy. During training, two subjects, A and B, are randomly selected. Given the same emotional stimulus, their corresponding EEG signal segments, ${X_A}$ and ${X_B}$, are extracted from the same time interval and constructed as a positive sample pair. This strategy effectively captures latent cross-subject semantic consistency without relying on additional data augmentation.

\subsubsection{A Feature Encoder with Dynamic Retention Time}
This paper adopts the encoder architecture of DAEST \cite{shen2025dynamic} and follows a processing pipeline of "temporal modeling–multiscale spatiotemporal feature extraction–attention enhancement–feature projection." First, temporal convolution is applied to the input EEG segments to extract local temporal patterns. Then, four parallel spatial transition convolution branches with different dilation rates are used to jointly model short-term and long-term dependencies, thereby capturing cross-channel coupling characteristics. A channel attention mechanism is further introduced to the concatenated multiscale features to adaptively reweight responses across different scales and channels, enhancing emotion-discriminative information. After the encoding process, the resulting features are unfolded along the temporal dimension into a sequence representation composed of ${T'}$ temporal tokens. It should be noted that each token corresponds to a local receptive field on the original timeline rather than a single sampling point, thus preserving rich local temporal dynamics at the sequence level.

During the contrastive learning phase, the model uses an encoder with shared parameters to map paired samples into a unified feature space.

\subsection{Asynchronous Contrastive Loss: Async-InfoNCE}
\subsubsection{Asynchronous Soft Matching and Maximum Similarity Calculation}
For a pair of EEG signals processed by the encoder described above, we define the feature sequence of the anchor sample as $U$ and the feature sequence of the comparison sample as $V$:
\begin{equation}
\begin{aligned}
    U &= \left[u_1,u_2,\ldots,u_{T_u}\right]^T \in \mathbb{R}^{T_u \times D}, \\
    V &= \left[v_1,v_2,\ldots,v_{T_v}\right]^T \in \mathbb{R}^{T_v \times D}.
\end{aligned}
\end{equation}
Here, ${D}$ denotes the feature dimension, and $T_u$ and $T_v$ represent the temporal lengths of sequences ${U}$ and ${V}$, respectively.

EEG signals exhibit significant temporal variability and non-stationarity. To address the challenge of achieving strict temporal alignment due to latency variations, this paper introduces a cross-sequence local matching mechanism in the temporal dimension. The traditional MaxSim similarity calculation strategy in NLP , involves calculating the similarity between tokens in the query and document samples separately, taking the maximum value as the best-match score for each token in the query sequence within the document sequence, and then summing these scores to obtain the overall similarity between the two samples.

\begin{equation}
    S_{q,d}=\sum_{i=1}^{\left|E_q\right|}{\max_{j=1}^{\left|E_d\right|}{E_{q_i}}\cdot E_{d_j}^T}
\end{equation}
Here, ${E_q=\{E_{q_i}\}_{i=1}^{\left|E_q\right|}}$ denotes the sequence of token representations for the query sample ${q}$, and ${E_d=\{E_{d_j}\}_{j=1}^{\left|E_d\right|}}$ denotes the sequence of token representations for the candidate sample $d$; ${E_{q_i}}$ and ${E_{d_j}}$ represent the $i$-th token representation in the query sequence and the $j$-th token representation in the candidate sequence, respectively; ${\left|E_q\right|}$ and ${\left|E_d\right|}$ represent the number of tokens in the two sequences, respectively. ${E_{q_i}\cdot E_{d_j}^T}$ denotes the dot-product similarity between the two token representations. This method achieves token-level local matching by finding the most similar token in the candidate sequence for each query token.

Basing on the MaxSim similarity calculation, we further introduce a ${TopK}$ retrieval mechanism. This ensures that the model no longer relies solely on a single strongest response but instead comprehensively considers multiple high-response local match results, thereby capturing a richer set of physiological cues related to potential emotions. Specifically, for each time-slice token ${u_i}$ in the anchor sequence ${U}$, we calculate its similarity with all time-slice tokens in sequence ${V}$ and construct a similarity set:
\begin{equation}
    S\left(u_i,V\right)=\{u_i^Tv_j\mid1\le j\le T_v\}
\end{equation}
Next, the similarity values in this set are sorted in descending order, and the top ${K}$ largest values are selected, denoted as:
\begin{equation}
    TopK\left(u_i,V\right)=\left[s_i^{\left(1\right)},s_i^{\left(2\right)},\ldots,s_i^{\left(K\right)}\right]
\end{equation}
This mechanism achieves "soft alignment" in the temporal dimension by allowing anchor signals to match similar responses across multiple time slices in a contrastive sequence.

\subsubsection{Local Similarity Aggregation}
After the matching calculation is completed for all time-slice features in the anchor sequence $U$, the corresponding local ${TopK}$ similarity scores are obtained. These local similarity scores are then averaged to define the asymmetric global similarity ${S_{U\rightarrow V}}$ from sequence ${U}$ to sequence ${V}$.

Specifically, for each time-slice feature ${u_i}$, the ${K}$ corresponding local optimal matching scores are denoted as ${\{s_i^{\left(k\right)}\}_{k=1}^K}$. By taking the arithmetic mean of these scores, we obtain the similarity of the time-slice feature ${u_i}$. In the process of aggregating global similarity, this paper adopts a mean aggregation method instead of the summation operation used in MaxSim. This approach controls the numerical scale of the similarity scores, thereby preventing gradient instability caused by excessively large values during subsequent contrastive learning optimization and improving the stability of model training.
\begin{equation}
    S_{U\rightarrow V}=\frac{1}{T_u}\sum_{i=1}^{T_u}\left(\frac{1}{K}\sum_{k=1}^{K}s_i^{\left(k\right)}\right)
\end{equation}
\subsubsection{Building the Async-InfoNCE Loss Function}
In the learning framework, given an anchor sequence $U$, a positive sample sequence ${V^+}$, and ${M}$ negative sample sequences ${{\{V_m^-}\}_{m=1}^M}$, the asynchronous similarity is defined as ${S_{async}(U,V){=S}_{U\rightarrow V}}$. Based on this, we incorporate the aforementioned asynchronous similarity metric into the traditional InfoNCE form to obtain the final Async-InfoNCE loss function:
\begin{equation}
    L_{Async}=-\log{\frac{\exp(S_{async}(U,V^+)/\tau)}{\exp{\left(S_{async}\left(U,V^+\right)/\tau\right)}+\sum_{m=1}^{M}{\exp(S_{async}(U,V^-)/\tau)}}}
\end{equation}
Here, ${\tau}$ is a temperature hyperparameter used to control the model's penalty for difficult negative samples. By minimizing ${L_{Async}}$, the model can effectively pull representations with similar emotional patterns across participants closer together while preserving fine-grained temporal dynamics, and push apart representations with dissimilar emotional patterns, thereby alleviating the issue of temporal asynchrony in EEG signals.

\subsection{Cross-subject emotion recognition}
After completing pre-training based on Async-InfoNCE, the spatio-temporal encoder ${f_\theta(\cdot)}$ has already demonstrated strong cross-subject representation capabilities. In the downstream sentiment classification task, we no longer use the Projector from the contrastive learning phase, but instead treat the pre-trained encoder as a fixed feature extractor. After loading the pre-trained parameters, the model outputs the attention-weighted intermediate temporal feature tensor ${z_{cls}}$ in a forward prediction mode.

Following feature extraction, to further mitigate cross-subject distribution shifts and temporal fluctuations, we apply normalization and LDS smoothing to the offline features in sequence, and then construct the training or validation set for input to the classification network. The classifier employs a multi-layer perceptron ${g_\phi(\cdot)}$, which is supervised and optimized via cross-entropy loss, and outputs category probabilities as:
\begin{equation}
    \hat{y}=Softmax(g_\phi(z_{cls}))
\end{equation}
At this stage, the encoder no longer participates in the parameter updates via backpropagation. Only the classifier parameters ${\phi}$ are trained. This decoupled paradigm, consisting of a pre-trained encoder and a lightweight classification head, preserves asynchronously aligned temporal information while reducing the interference of the projection head in the downstream discriminative space, thereby improving the stability and generalization performance of cross-subject emotion recognition.

\begin{figure}[t]
\centering
\includegraphics[width=0.9\columnwidth]{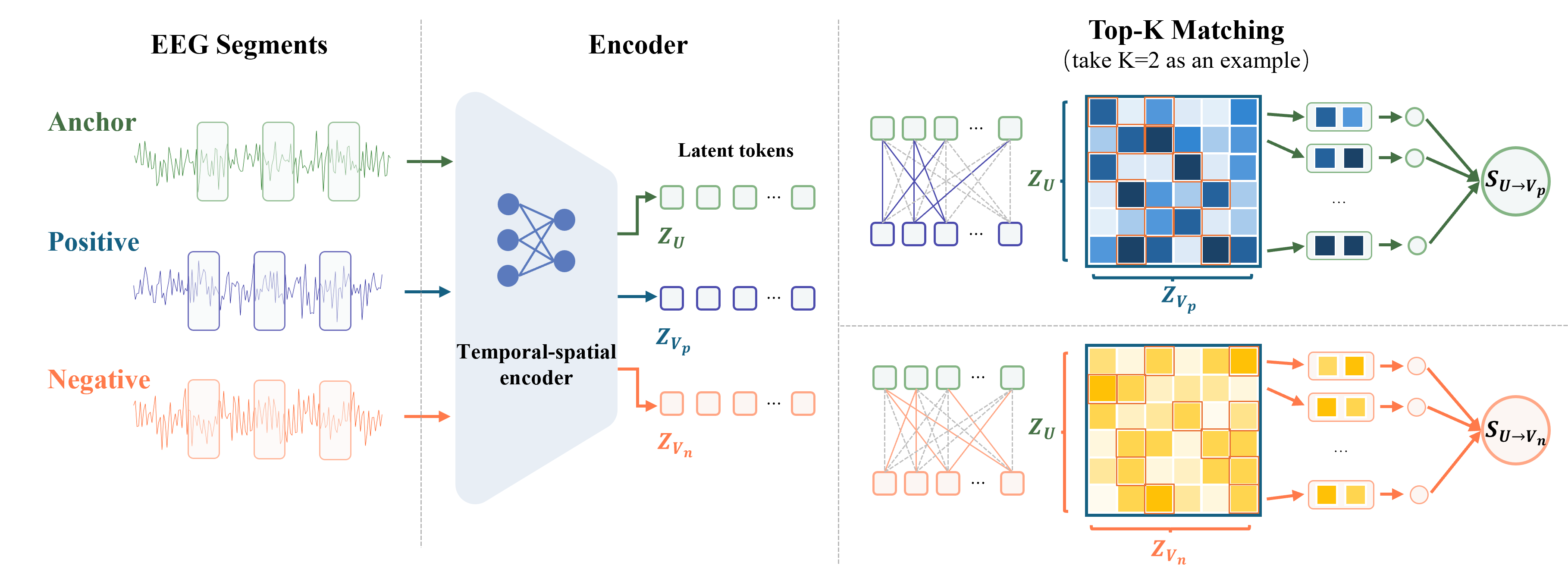}
\caption{Schematic diagram of Async-InfoNCE based on temporal token matching. Given an anchor sequence $U$ and contrastive sequence $V$, the encoded EEG features are reorganized along the temporal dimension into a temporal token representations, and a pairwise similarity matrix is computed between the two sets of tokens. For each token in anchor sequence, the $K$ tokens with the highest similarity are selected from the comparison sequence, and their mean is calculated to obtain the row-wise asynchronous matching score.}
\label{method}
\end{figure}

\section{Experiments}
\label{sec:experiments}

\subsection{Dataset}
We validated our model on three datasets: FACED \cite{chen2023large}, SEED \cite{zheng2015investigating,duan2013differential}, and SEED-V \cite{liu2021comparing}. SEED is a three-class dataset and is one of the most widely used benchmarks in EEG-based emotion recognition. In contrast, FACED contains nine emotion categories, while SEED-V contains five categories, making them more suitable for fine-grained analysis and discussion.

FACED is a large-scale, fine-grained emotional EEG dataset released by Tsinghua University, representing a significant breakthrough in terms of sample size and classification level. The dataset comprehensively records 32-channel EEG signals from up to 123 participants while they watched 28 emotion-eliciting videos, with a raw sampling rate of 250 Hz. In the binary classification task, labels are categorized into positive and negative emotions, while in the nine-class classification task, they are further refined into joy, inspiration, tenderness, and happiness (positive), anger, fear, disgust, and sadness (negative), as well as neutral emotions. The inclusion of the FACED dataset provides an important basis for evaluating the proposed method. On the one hand, its nine-class emotion labels are well suited to assessing the model's ability to distinguish fine-grained emotional states. On the other hand, the relatively large number of participants, namely 123 subjects, introduces substantial inter-subject physiological variability and distribution shifts, providing a challenging evaluation setting for cross-subject emotion recognition. This enables a comprehensive assessment of the model's ability to learn subject-invariant representations and achieve robust generalization across subjects.

The SEED dataset, constructed and released by a team at Shanghai Jiao Tong University, is one of the most widely used public datasets in the field of EEG-based affective computing. The dataset recruited 15 participants (including 7 males and 8 females). To record changes in emotional states over time, each participant was asked to complete three separate EEG recording sessions on different dates, for a total of 45 experimental sessions. In the experimental design for each session, participants were asked to watch 15 film clips. Each video clip lasted approximately 4 minutes and was intended to elicit one of three emotional states: positive, negative, or neutral. The experimental paradigm consisted of video playback, self-report assessment, and rest periods. During this process, researchers used a 62-channel ESI NeuroScan system to collect real-time scalp EEG signals from the participants, with the device's raw data sampling rate set to 1000 Hz. In the experimental setup described in this paper, we conducted a cross-participant three-class EEG emotion recognition experiment based on the EEG signals from this dataset.

SEED-V, also collected and publicly released by a team at Shanghai Jiao Tong University, is a multimodal dataset built upon the original SEED dataset to capture more complex emotional dimensions. The project recruited a total of 20 participants for data collection and adopted a cross-session experimental design similar to SEED, in which participants were required to complete three independent sessions on different dates. During each session, participants were asked to watch 15 specific video clips. The system recorded their EEG activity throughout the viewing process using a 62-channel EEG device. Additionally, to support multimodal data analysis, SEED-V simultaneously collected participants' eye-tracking data while recording EEG signals. SEED-V's labeling system expands the emotional categories to five: happy, sad, fearful, disgusted, and neutral. This classification significantly refines the internal boundaries of negative emotions (sadness, fear, and disgust). In the experimental section of this paper, we utilized the EEG data provided by this dataset to conduct a cross-subject five-class EEG emotion recognition experiment.

\subsection{Data Preprocessing}
Our preprocessing method follows the DAEST model \cite{shen2025dynamic}, applying the same preprocessing to the FACED, SEED, and SEED-V datasets. First, the signals are downsampled to 125 Hz and passed through a bandpass filter with a bandwidth of 0.5–47 Hz; then, the signals are divided into independent samples based on the start times of the emotional video stimuli. To eliminate noise interference, a dual-threshold detection mechanism is implemented to filter out noisy channels, where $m$ represents the amplitude threshold and $n$ represents the noise duration threshold. Two sets of thresholds were used to filter out noise: $m$ = 3 and $n$ = 0.4 to identify long-duration artifacts, and $m$ = 30 and $n$ = 0.01 to identify short-duration, high-amplitude artifacts. The identified noise channels were repaired using interpolation. Next, spurious signals caused by eye movements and active muscle activity are removed using independent component analysis (ICA). Components related to eye and muscle activity with a confidence score exceeding 0.8 are removed using ECGLab.

For the FACED dataset, 10-fold cross-validation was employed, dividing all participant data into ten groups, with nine groups (each containing 12 participants) serving as the training set and the final group (containing 15 participants) serving as the test set; For the SEED dataset, a five-fold cross-validation protocol was used, dividing all participant data into five groups, with three participants in each group; for the SEED-V dataset, the leave-one-out method was used, whereby the sample from one participant was selected as the test set each time, and the samples from the remaining participants formed the training set.

\subsection{Implementation Details}
For pre-training, we used a learning rate of 0.0007, a weight decay of 0.00015, and 10 training iterations. For classifier training, we used a learning rate of 0.005 and a weight decay of 0.0022, with the maximum and minimum numbers of training iterations set to 100 and 30, respectively. The early stopping patience was set to 30. Other key hyperparameters used for different datasets are summarized in Table~\ref{tab:Hyperparameter}.

\begin{table}
\centering
\caption{Hyperparameter Settings for Different Datasets}
\label{tab:Hyperparameter} 
\scriptsize
\begin{tabular}{ccccc}
\toprule
Hyperparameter & SEED(cls3) & SEEDV(cls5) & FACED(cls2) & FACED(cls9)\\
\midrule
n\_timeFilters & 16 & 16 & 16 & 16 \\
timeFilterLen & 30 & 30 & 30 & 30 \\
n\_msFilters & 4 & 4 & 4 & 4 \\
msFilter\_timeLen & 3 & 3 & 3 & 3 \\
dilation\_array & [1,3,6,12] & [1,3,6,12] & [1,3,6,12] & [1,6,12,24] \\
avgPoolLen & 15 & 15 & 15 & 15 \\
timeSmootherLen & 3 & 3 & 3 & 3 \\
Weight Decay & 0.015  & 0.015  & 0.015  & 0.015 \\
Dropout & 0.1 & 0.1 & 0.1 & 0.1 \\
\bottomrule
\end{tabular}
\end{table}

\section{RESULTS AND ANALYSES}
\subsection{Hyperparameter Analysis}
In this paper, the similarity aggregation of time-slice tokens employs a ${TopK}$ matching strategy. To reduce search complexity, the search is limited to the range ${K \in \{1, 2, 3\}}$. The hyperparameter $K$ controls the number of local responses matched by each time-slice token within the candidate sequences. When ${K}$ = 1, the model matches only with the token in the candidate sequence that has the highest similarity, highlighting the strongest local correspondence. However, this approach is susceptible to interference from random peaks or local noise, potentially overlooking other candidate tokens that are equally highly correlated. As ${K}$ increases, the model integrates the matching results of multiple high-response tokens, making the asynchronous similarity estimation more robust and better able to withstand temporal shifts and local noise. However, this comes at the cost of potentially weakening the model's ability to distinguish the strongest local matches, leading to blurred similarity boundaries between positive and negative samples.

\begin{figure}[htbp]
\centering
\includegraphics[width=0.6\columnwidth]{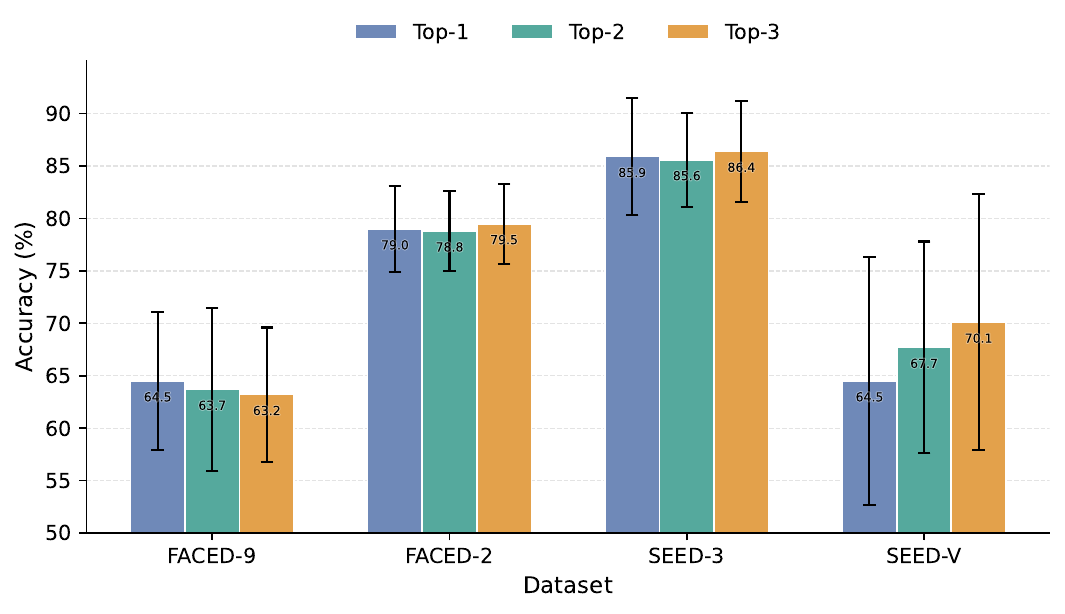}
\caption{The performance of different K values across various datasets}
\label{fig:topk}
\end{figure}

As shown in Table \ref{tab:TopK} and Figure \ref{fig:topk}, the optimal ${K}$ values for different datasets are not entirely consistent. In the FACED nine-class classification task, ${K}$=1 achieved the highest accuracy, indicating that in scenarios with a large number of categories and finer-grained emotional classifications, the strongest local matches are sufficient to carry the primary discriminative information, while introducing more tokens may instead lead to redundant matches or confusion between categories. In contrast, for the FACED binary classification, SEED three-class, and SEED V five-class tasks, $K$=3 yields the best results. This suggests that in these tasks, aggregating multiple high-similarity local fragments leads to more stable similarity estimates.

\begin{table}[htbp]
\centering
\caption{Hyperparameter Settings for Different Datasets}
\label{tab:TopK} 
\scriptsize
\begin{tabular}{ccc}
\toprule
Dataset & ${TopK}$ Value & Acc(\%)\\
\midrule
FACED(cls9) & 1 & \textbf{64.5±6.6} \\
FACED(cls9) & 2 & 63.7±7.8  \\
FACED(cls9) & 3 & 63.2±6.4 \\
\midrule
FACED(cls2) & 1 & 79.0±4.1 \\
FACED(cls2) & 2 & 78.8±3.8 \\
FACED(cls2) & 3 & \textbf{79.5±3.8} \\
\midrule
SEED(cls3) & 1 & 85.9±5.6 \\
SEED(cls3) & 2 & 85.6±4.5 \\
SEED(cls3) & 3 & \textbf{86.4±4.8} \\
\midrule
SEEDV(cls5) & 1 & 64.5±11.8 \\
SEEDV(cls5) & 2 & 67.7±10.1 \\
SEEDV(cls5) & 3 & \textbf{70.1±12.2} \\
\bottomrule
\end{tabular}
\end{table}

To further validate the validity of the $TopK$ matching mechanism, we calculated the similarity variance of the ${Top3\left(u_i,V\right)}$ for each time-slice token (i.e., ${u_i}$) on the SEED dataset for both ${K}$=1 and ${K}$=3. Since the experiment employed five-fold cross-validation, we selected the folds where ${K}$=1 outperformed ${K}$=3 and the folds where ${K}$=3 outperformed ${K}$=1 from the validation set results to compare the two scenarios.

\begin{figure}[h]
\centering
\includegraphics[width=0.6\columnwidth]{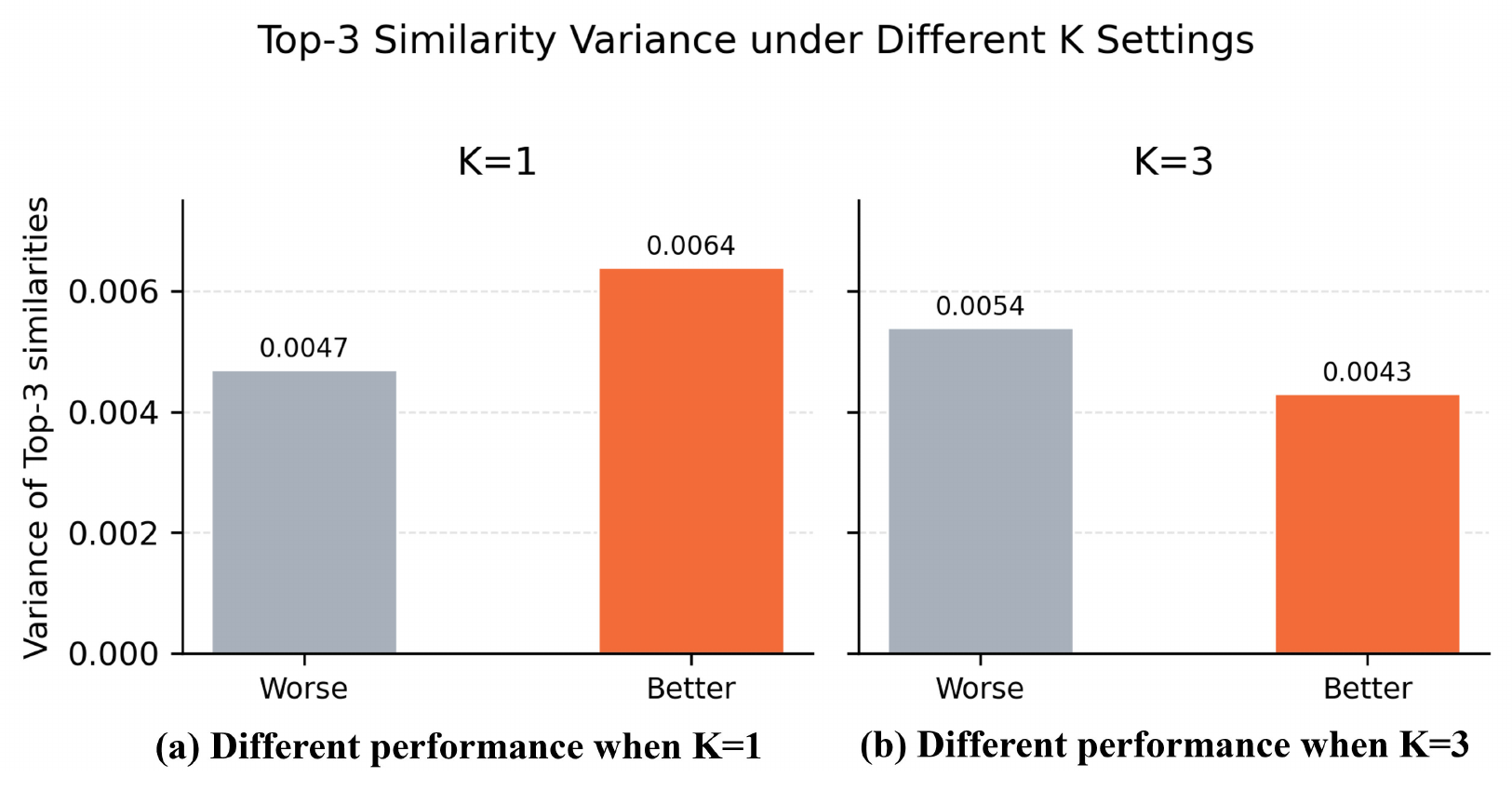}
\caption{Analysis of variance for $Top3$ token similarity under different $K$ settings. The figure compares the variance of ${Top3\left(u_i,V\right)}$ pairs for $K$=1 and $K$=3 under both poor and good performance conditions. (a) A larger variance indicates more pronounced differences among high-ranking matches, suggesting that additional matches may introduce unstable information, making a smaller $K$ value more appropriate. (b) A smaller variance indicates greater consistency among multiple highly similar matches, making a larger $K$ value more suitable for aggregation.}
\label{fig:variance}
\end{figure}

As shown in Figure \ref{fig:variance}, when ${K}$=1 yields better results, the variance of ${Top3\left(u_i,V\right)}$ is relatively larger, indicating that, apart from the match with the highest similarity, there are significant differences between the other high-ranking matches and the optimal match. In this case, introducing more candidate matches may incorporate locally similar but weakly correlated or unstable segments into the similarity calculation, thereby weakening the discriminative power of the strongest match and degrading model performance. In contrast, when ${K}$=3 yields better results, the variance of ${Top3\left(u_i,V\right)}$ is relatively smaller. This indicates that the similarity among multiple high-ranking matches is closer, suggesting that additional candidate matches are not merely noise but are more likely to contain valid local correspondence information. Therefore, in such cases, retaining only the highest-similarity match may result in the loss of useful local matching information, whereas using ${K}$=3 enables a more comprehensive and stable estimate of asynchronous similarity.

The above results demonstrate that the ${TopK}$ matching mechanism can strike a balance between a single strongest match and multiple high-response matches based on differences in the local matching distribution. Compared to the MaxSim strategy, which relies solely on the maximum response, this mechanism integrates multiple potentially valid local matching results, thereby providing richer information for modeling emotion-related temporal segments.

\subsection{Performance Comparison}
To evaluate the effectiveness of the proposed TA2CL, we compare its performance with a baseline method and several state-of-the-art methods on cross-subject EEG emotion recognition tasks. The baseline method, DE+MLP, first extracts Differential Entropy (DE) features from EEG signals and then uses a Multi-Layer Perceptron (MLP) for classification. In addition, we compare TA2CL with several representative methods, including DS-AGC \cite{wehrmann2018hierarchical}, DAEST \cite{shen2025dynamic}, GCPL \cite{russell1980circumplex}, PPDA \cite{zhao2021plug}, and CLISA \cite{shen2022contrastive}. DS-AGC combines self-attention-based adversarial learning with graph structures to capture complex relationships among subjects in EEG data. DAEST models the dynamic spatiotemporal characteristics of EEG signals through attention mechanisms. GCPL integrates contrastive learning with domain generalization to learn features that are both class-discriminative and domain-invariant. PPDA employs shared and private encoders to jointly capture subject-invariant and subject-specific features, while CLISA improves cross-subject generalization by learning subject-invariant representations.

To ensure experimental comparability, all methods were evaluated under a fair benchmark setting. Overall, the performance on multi-class classification tasks is generally lower than that on binary or lower-class classification tasks, indicating that fine-grained emotion classification remains challenging. Meanwhile, the performance on FACED is generally lower than that on the SEED and SEED-V, which may be attributed to its more fine-grained emotion categories and greater cross-subject variability.

\begin{figure}
    \centering
    \includegraphics[width=0.8\linewidth]{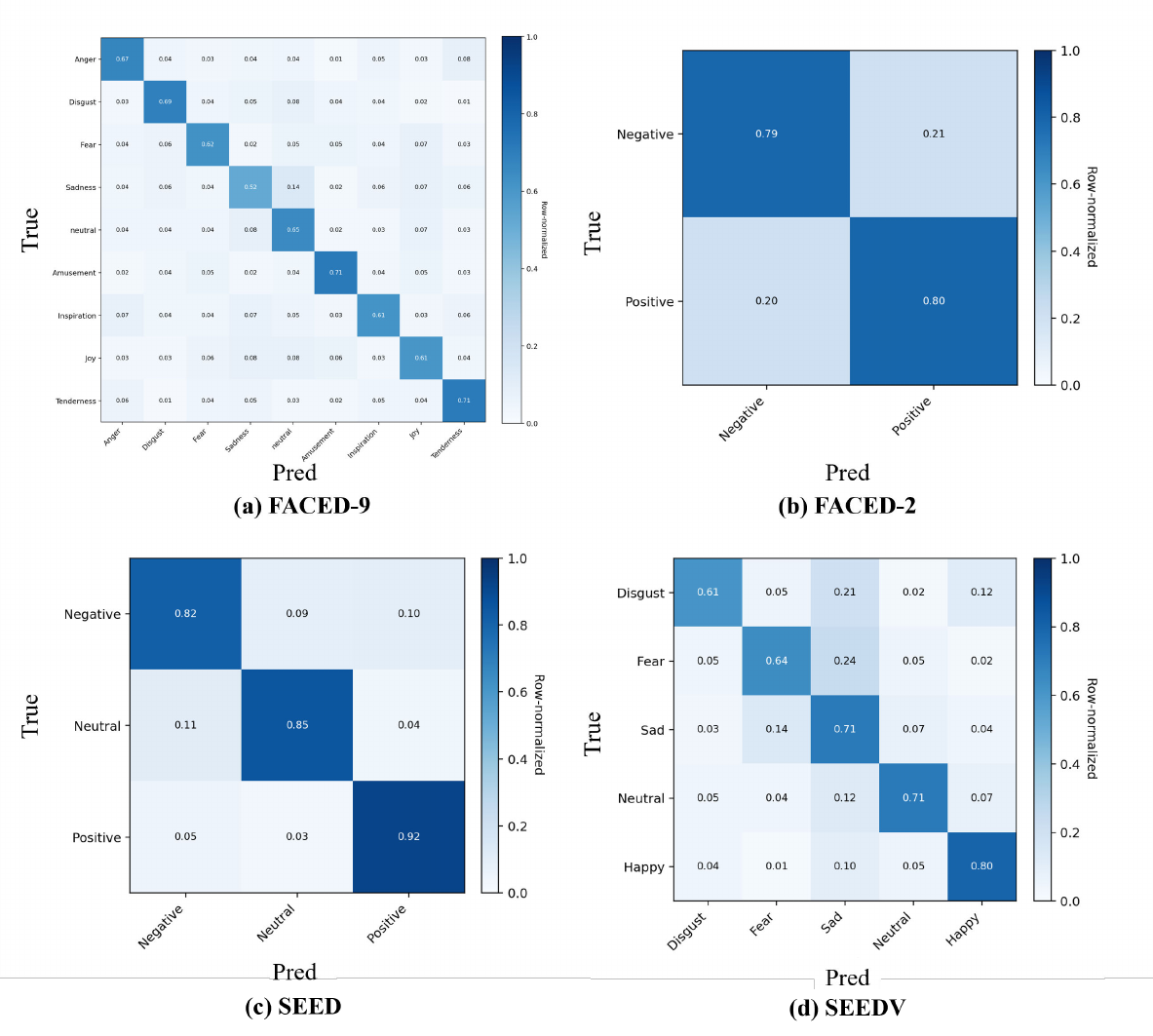}
    \caption{Confusion matrices of the datasets FACED, SEED, and SEEDV}
    \label{fig:matrix}
\end{figure}

The proposed TA2CL achieves superior performance compared with existing methods in most experimental settings. On the FACED dataset, TA2CL achieves an accuracy of 79.5\%±3.8\% in the binary classification task and 64.5\%±6.6\% in the nine-class classification task, showing a clear improvement over DAEST. On the SEED dataset, TA2CL achieves an accuracy of 86.4\%±4.8\%, while on the SEED-V dataset, it achieves 70.1\%±12.2\%. Detailed results are presented in Table \ref{tab:FACED} and Table \ref{tab:SEED}, and the corresponding confusion matrices are shown in Figure \ref{fig:matrix}.

\begin{table}[t]
\centering

\begin{minipage}{0.45\linewidth}
\centering
\caption{Cross-subject emotion recognition performance on the FACED dataset}
\label{tab:FACED} 
\scriptsize
\begin{tabular}{cccc}
\toprule
Methods	& FACED-2 (Acc\%) & FACED-9 (Acc\%)	& Source of Results\\
\midrule
DE+MLP & 60.9±3.2 & 21.7±2.2 & Baseline \\
TCA\cite{zheng2015transfer} & / & 36.0±4.1 & Original Results \\
DS-AGC\cite{wehrmann2018hierarchical} & / & 15.1±0.3 & Original Results  \\
GCPL\cite{russell1980circumplex} & / & 36.9±3.3 & Original Results \\
SimCLR\cite{chen2020simple} & / & 30.2±2.7 & Original Results \\
CLISA\cite{shen2022contrastive} & 67.8±4.1 & 42.4±3.1 & Reproduced Results \\
DAEST\cite{shen2025dynamic} & 78.3±4.8 & 62.4±8.1 & Reproduced Results \\
\textbf{ours} & \textbf{79.5±3.8} & \textbf{64.5±6.6}	\\
\bottomrule
\end{tabular}
\end{minipage}
\hfill
\begin{minipage}{0.45\linewidth}
\centering
\caption{Cross-subject emotion recognition performance on the SEED dataset}
\label{tab:SEED} 
\scriptsize
\begin{tabular}{cccc}
\toprule
Methods	& SEED (Acc\%)	& SEED-V (Acc\%) & Source of Results\\
\midrule
DE+MLP	& 79.9±8.7 & 59.3±17.2 & Baseline \\
TCA\cite{zheng2015transfer}	& 76.7±7.0 & / & Original Results \\
GCPL\cite{russell1980circumplex} & 80.7±6.0 & / & Original Results \\
SimCLR\cite{chen2020simple} & 68.4±7.7 & / & Original Results \\
CLISA\cite{shen2022contrastive} & 86.4±6.4 & 67.3±13.0 & Reproduced Results \\
PPDA\cite{zhao2021plug} & 86.7±7.1 & / & Original Results \\
DAEST\cite{shen2025dynamic} & 86.1±6.2 & 68.2±12.4 & Reproduced Results \\
\textbf{ours} & \textbf{86.4±4.8} & \textbf{70.1±12.2} \\
\bottomrule
\end{tabular}
\end{minipage}

\end{table}

\subsection{Ablation Study}
To evaluate the role of the asynchronous contrastive learning module in this model, we conducted several ablation experiments to investigate its overall impact on the model.
\subsubsection{Ablation of the attention mechanism}
To further analyze the synergy between the attention mechanism and asynchronous contrastive learning, this paper conducts ablation experiments on the attention module under identical training conditions on the FACED-9 and SEED datasets, and compares the proposed TA2CL with DAEST. The experimental results are shown in Table \ref{tab:att}. On the FACED-9 dataset, the accuracy of DAEST was 49.4\% after removing the attention module, and it increased to 62.4\% after adding the attention mechanism, representing a 13.0\% improvement; meanwhile, the accuracy of TA2CL was 48.4\% without attention, and it increased to 64.5\% after adding the attention module, representing a 16.1\% improvement. On the SEED dataset, DAEST improved from 85.8\% to 86.1\% after adding attention, a 0.3\% increase; TA2CL improved from 85.2\% to 86.4\%, a 1.2\% increase. It can be seen that on both datasets, TA2CL achieved greater performance gains after introducing the attention mechanism, indicating a certain degree of synergy between asynchronous contrastive learning and the attention module. Specifically, the attention module can adaptively weight the responses of different feature channels based on local temporal context, thereby highlighting emotion-related temporal patterns; meanwhile, the asynchronous contrastive loss further models the non-strict alignment relationships between local segments across samples at the token level.

\begin{figure}[h]
    \centering
    \includegraphics[width=0.5\linewidth]{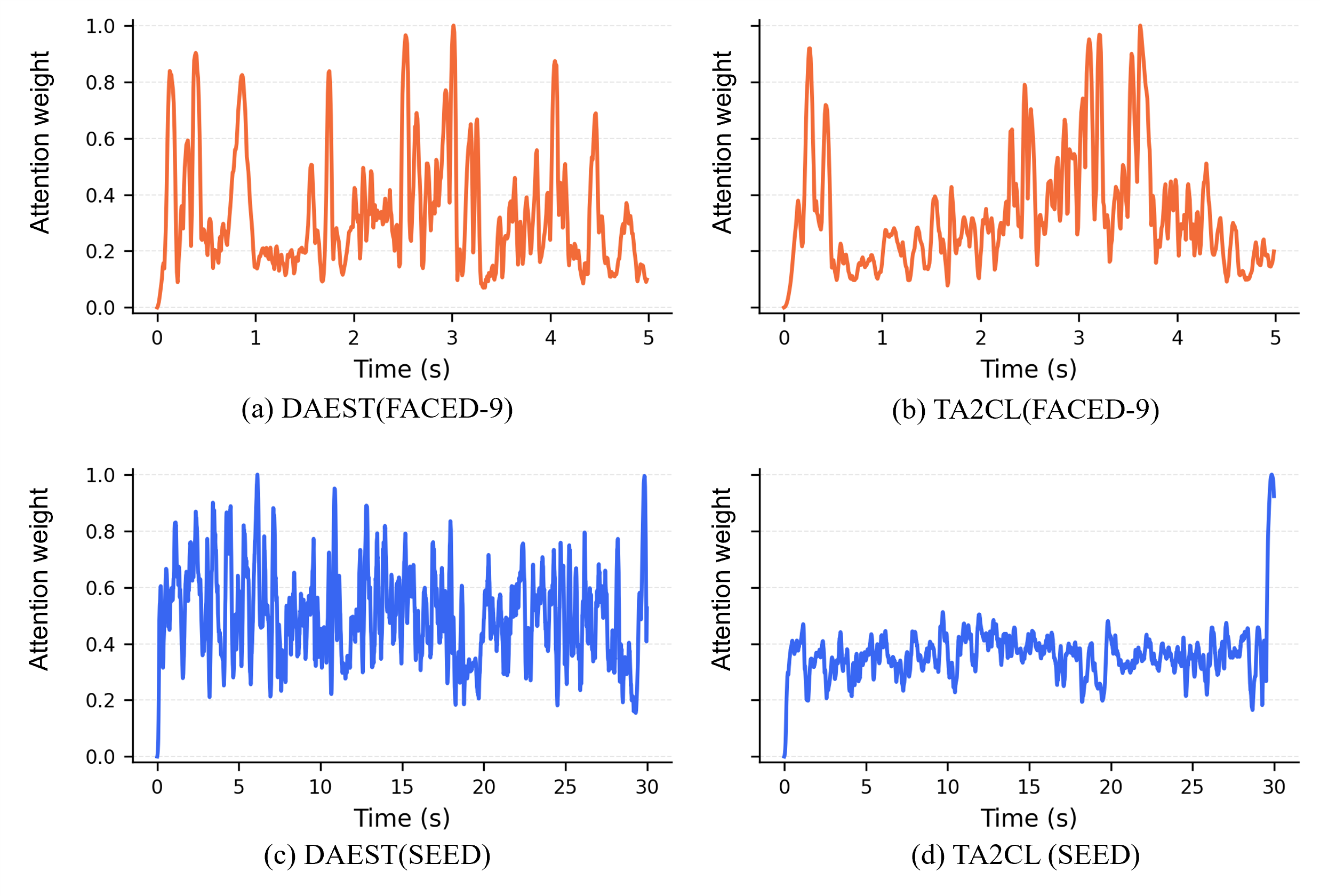}
    \caption{Visualization of attention responses of DAEST and TA2CL on FACED-9 and SEED.The curves are obtained by max aggregation of channel attention weights along the channel dimension, reflecting the strongest attention response at each temporal position. Compared with DAEST, TA2CL exhibits more continuous high-response regions, indicating its ability to capture more stable emotion-related temporal patterns.}
    \label{fig:att}
\end{figure}

Figure \ref{fig:att} shows the visualizations of attention weights for DAEST and TA2CL on the FACED-9 and SEED datasets. The attention responses in the figure are obtained by aggregating channel attention weights along the temporal dimension. Compared to DAEST, TA2CL exhibits more continuous high-response regions within local time intervals, whereas DAEST's attention responses manifest primarily as scattered transient peaks. This indicates that the proposed asynchronous contrastive learning strategy helps the attention module capture more stable local temporal patterns, thereby enhancing the representation capacity of emotion-related features. Consequently, the combination of the two approaches can more effectively mitigate temporal asynchrony in EEG signals and improve the discriminative ability of emotion representation. It is worth noting that, compared to FACED-9, the absolute improvement brought by the attention module on SEED is smaller. This may be attributed to the smaller number of categories in the SEED dataset and its already high overall classification performance. Under these circumstances, the room for improvement from the attention mechanism is relatively limited; however, TA2CL still demonstrates a more pronounced trend of improvement.

\begin{table}[h]
\centering
\caption{Ablation study of the attention module on FACED-9 and SEED}
\label{tab:att} 
\scriptsize
\begin{tabular}{ccccc}
\toprule
Dataset & Method & w/o Attention Acc (\%) & w/ AttentionAcc (\%) & Gain (\%) \\
\midrule
FACED-9 & DAEST & 49.4±6.6 & 62.4±8.14 & +13.0 \\
FACED-9 & TA2CL & 48.4±6.4 & 64.5±6.6 & +16.1 \\
\midrule
SEED & DAEST & 85.8±4.7 & 86.1±6.2 & +0.3 \\
SEED & TA2CL & 85.2±4.6 & 86.4±4.8 & +1.2 \\
\bottomrule
\end{tabular}
\end{table}

\subsubsection{The Impact of Different Aggregation Strategies on Results}
Furthermore, to investigate how different aggregation strategies in Async-InfoNCE affect model performance, we conduct ablation experiments at two levels. The first level concerns the aggregation of the ${TopK}$ similarities for each anchor token, denoted as ${TopK}$ Aggregation. The second level concerns the integration of local matching scores from all anchor tokens into a global similarity score, denoted as Token Aggregation. In the default setting, mean aggregation is first applied to the $TopK$ similarity scores of each anchor token to obtain its local asynchronous matching score. The local matching scores of all anchor tokens are then averaged to compute the final global similarity score.

To examine the impact of different token aggregation methods, we compare three strategies: Mean, Sum, and Weighted Mean. Mean aggregation normalizes the contribution of tokens involved in the matching process, thereby keeping the similarity scale stable. Sum aggregation directly accumulates the matching scores, which preserves a stronger overall response but may lead to excessively large similarity values as the number of tokens or the ${TopK}$ count increases. Weighted Mean, in contrast, introduces an additional MLP to learn adaptive weights for different tokens, enabling more informative tokens to contribute more prominently to the final similarity score. This comparison further investigates whether weighted aggregation can improve the modeling capacity of local matching.

The experimental results are shown in Table \ref{tab:aggregation}. When $K$=1, token-level Mean aggregation achieves accuracies of 64.5\% and 79.0\% on FACED-9 and FACED-2, respectively, outperforming token-level Sum aggregation, which achieves 45.3\% and 70.3\%. This result suggests that directly summing the local matching scores of all anchor tokens does not necessarily lead to better performance in global similarity computation. A possible reason is that sum aggregation causes the final similarity score to increase significantly as the number of tokens accumulates, leading to excessively large logits in the InfoNCE loss. When the temperature parameter ${\tau}$ is fixed, overly large similarity values may make the softmax distribution excessively sharp, thereby weakening effective gradients and reducing training stability. In contrast, Mean aggregation provides normalization over the token dimension, keeping the similarity scores between different samples within a more stable numerical range. Meanwhile, weighted Mean aggregation achieves an accuracy of 63.1\% on FACED-9, which is slightly lower than Mean aggregation. On FACED-2, it achieves 79.0\%, which is essentially consistent with Mean aggregation. This indicates that introducing learnable token weights does not bring consistent performance gains. One possible explanation is that the ${TopK}$ matching mechanism in Async-InfoNCE has already selected highly responsive local fragments at the token level. Introducing additional MLP-based weights increases optimization complexity and may lead to overfitting or instability when the sample size is limited. Therefore, compared with more complex weighting strategies, simple Mean aggregation achieves a better balance between training stability and generalization performance.

Furthermore, this paper compares the Mean and Sum strategies in ${TopK}$ Aggregation. When ${K}$=3, the Mean aggregation within ${TopK}$ achieved accuracy rates of 63.2\% and 79.5\% on FACED-9 and FACED-2, respectively, both outperforming the 62.9\% and 78.2\% achieved by Sum aggregation. This indicates that for multiple highly similar matches for each anchor token, direct summation—while capable of enhancing local responses—may also amplify the impact of redundant or noisy matches; in contrast, Mean aggregation preserves information from multiple potentially valid local matches while avoiding an excessively large similarity scale, thereby yielding more stable local match scores.

In summary, Mean aggregation demonstrates more stable performance at both the token aggregation and $TopK$ aggregation levels. Therefore, this paper ultimately adopts the Mean-Mean aggregation method as the default setting for Async-InfoNCE: first, the Mean of the $TopK$ matching results for each anchor token is calculated; then, the Mean of the local matching scores for all tokens is calculated to obtain a globally asynchronous similarity that is scale-stable and robust.

\begin{table}
\centering
\caption{Ablation study on aggregation strategies in Async-InfoNCE}
\label{tab:aggregation} 
\scriptsize
\begin{tabular}{ccc}
\toprule
Method & FACED-9 & FACED-2 \\
\midrule
Token Aggregation(${K}$=1) \\
\midrule
Mean & 64.5±6.6 & 79.0±4.1 \\
Sum & 45.3±6.5 & 70.3±2.6 \\
Weighted Mean & 63.1±8.7 & 79.0±3.7 \\
\midrule
TopK Aggregation(${K}$=3)\\
\midrule
Mean & 63.2±6.4 & 79.5±3.8 \\
Sum & 62.9±7.3 & 78.2±4.0 \\
\bottomrule
\end{tabular}
\end{table}

\subsection{Visualization}
\begin{figure}[h]
    \centering
    \includegraphics[width=0.9\linewidth]{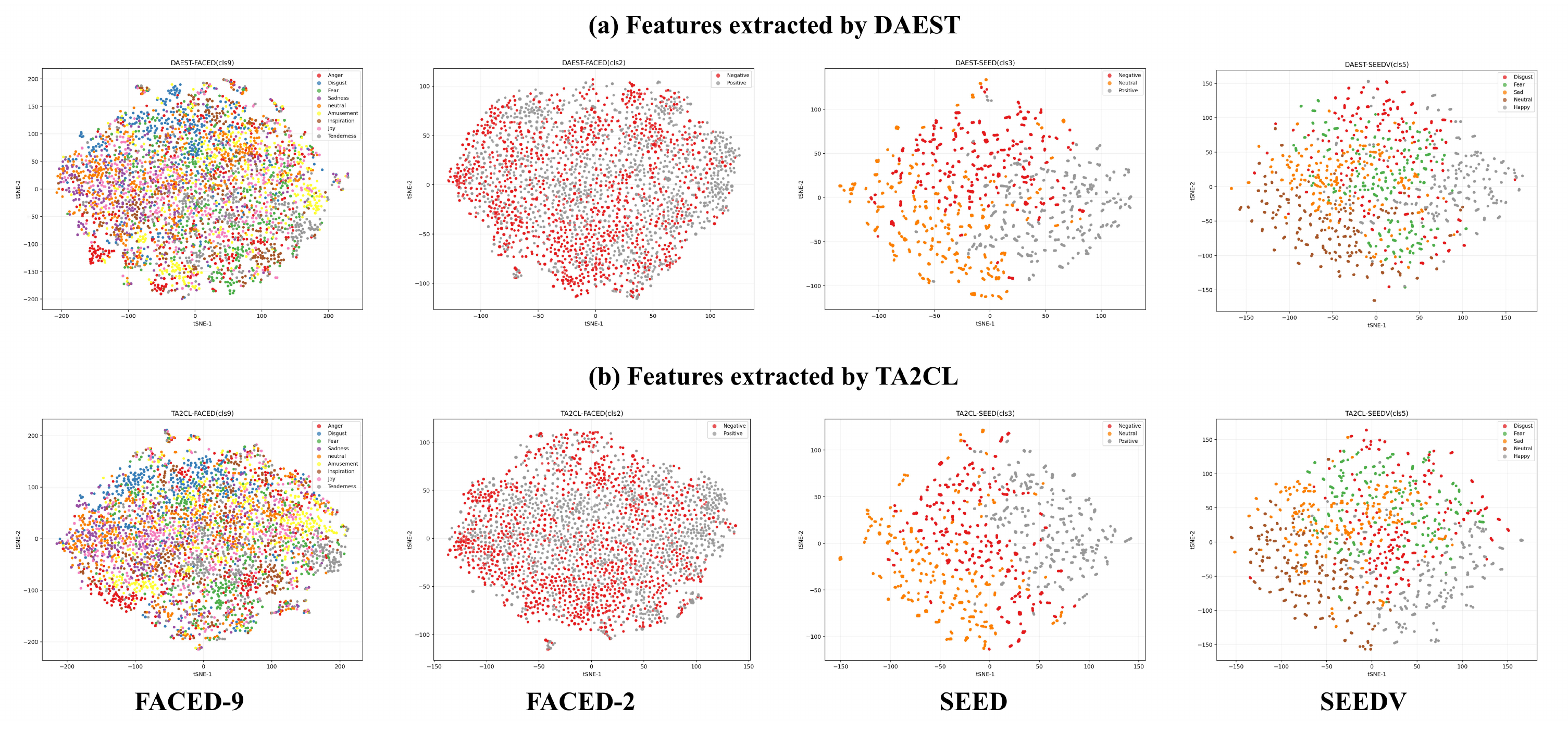}
    \caption{Comparison of tSNE plots for our model (TA2CL) and the DEAST model on the FACED, SEED, and SEEDV datasets}
    \label{fig:tSNE}
\end{figure}

Figure \ref{fig:tSNE} compares the feature distributions of the DAEST model and the TA2CL model proposed in this paper across four classification tasks, with different colors used to distinguish between emotion categories. It can be seen that, compared to DAEST, TA2CL exhibits clearer inter-class separability in the feature space, indicating that TA2CL possesses stronger discriminative capabilities in emotion feature learning.

However, as the classification is further refined, the feature distributions of different emotion categories begin to overlap more, and the model's ability to distinguish between emotions consequently declines. This phenomenon reflects that, in fine-grained emotion recognition scenarios, the differences between categories are inherently subtle. Coupled with the complexity introduced by inter-subject variations and the dynamic changes in emotions over time, this poses a greater challenge to the model's representational capabilities. Therefore, how to further enhance the model's discriminative power in fine-grained emotion recognition tasks remains a research direction worthy of continued exploration.

\section{Conclusion} 
\label{sec:conclusion}
Our study proposes a contrastive learning framework based on temporal asynchrony alignment, aimed at addressing the challenge of emotion-induced latency biases in cross-subject EEG signals. By uncovering and aligning latent emotional similarities across asynchronous time scales, this framework effectively extracts subject-invariant emotional representations. Extensive experimental results on three public EEG datasets demonstrate that the proposed method improves cross-subject emotion classification performance, fully validating its effectiveness and robustness. This study effectively addresses the shortcomings of existing models in handling temporal asynchrony in emotions and provides an innovative approach for learning generalizable features in future EEG-based emotion recognition. In future research, this framework is expected to be combined with network-driven deep decoding models and integrated into real-time brain-computer interface (BCI) closed-loop systems. By capturing and decoding the brain's spatiotemporal dynamic features in real time, it will facilitate the expansion of traditional static recognition into fields such as adaptive neural feedback and neurorehabilitation, thereby enabling more intuitive human-computer interaction and proactive mental health interventions \cite{forbes2026application,zeng2024adaptive}.

\bibliographystyle{IEEEtran}
\bibliography{main}

\end{document}